\documentclass[11pt]{article}
\usepackage{amsmath, amssymb}
\usepackage{amsbsy}
\usepackage{amsfonts}
\usepackage{amssymb}
\usepackage{amsmath}
\usepackage{graphicx}

\newcommand{\mathsym}[1]{{}}

\renewcommand{\title}[1]{\vbox{\center\LARGE{#1}}\vspace{5mm}}
\renewcommand{\author}[1]{\vbox{\center#1}\vspace{5mm}}

\def\ci{\cite}

\def \td {\tilde}

\def \N {{\mathcal N}}

\def \m {\mu}

\def \bi{\bibitem}
\def \la {\label}

\def \l {\lambda}
\def\foot{\footnote}

\def \sql {{\sqrt{\lambda}}}
\def \adss {$AdS_5 \times S^5$\ }
\newcommand{\rf}[1]{(\ref{#1})}
\def \ov {\over}

\def\th{\theta}

\def\N{{\cal N}}

\def \ha{{\textstyle{1\ov 2}}}

\def\r{{\rm r}}

\def \J {\mathcal{K}}
\def \del {\partial}

\def \J {{\cal J}}

\def \bi{\bibitem}
\def \la {\label}

\def \l {\lambda}
\def\foot{\footnote}

\def \sql {{\sqrt \l}}

\def \adss {$AdS_5 \times S^5$\ }

 \def \r {\rho}

\def \ov {\over}

\def \varpi {{\rm w}}

\def \no {\nonumber }

\def \adss {$AdS_5 \times S^5$\ }

\def \N {{\cal N}}

  \def \te {\theta}

\def \k {\kappa}
\def \ci {\cite}

\newcommand{\id}{\mathbf{1}}
\renewcommand{\imath}{i}

\renewcommand{\leq}{\,{\leqslant}\,}

\newcommand{\binner}[2]{%
  {\langle}\kern-4.15pt{\langle}#1{,}\,#2{\rangle}\kern-4.15pt{\rangle}}

\newcommand{\ffrac}[2]{\raisebox{.5pt}%
  {\footnotesize$\displaystyle\frac{#1}{#2}$}\kern1pt}

\def\id{\protect{{1 \kern-.28em {\rm l}}}}

\makeatletter
\renewcommand\section{\@startsection {section}{1}{\z@}%
                                   {-3.5ex \@plus -1ex \@minus -.2ex}%
                                   {2.3ex \@plus.2ex}%
                                   {\normalfont\large\bfseries}}
\renewcommand\subsection{\@startsection{subsection}{2}{\z@}%
                                   {-3.25ex\@plus -1ex \@minus -.2ex}%
                                   {1.5ex \@plus .2ex}%
                                   {\normalfont\normalsize\bfseries}}
\makeatother

\numberwithin{equation}{section} \makeatletter
\@addtoreset{equation}{section}

  \def \te {\theta}
  
  \def \a {\alpha}

\def \s {\sigma}

\newcommand{\be}{\begin{eqnarray}}
\newcommand{\ee}{\end{eqnarray}}

 \def \er {{\rm e}}

\def \bea{\be}
\def \eea{\ee}  
\def \edo {\end{document}}

\def \te {\textstyle}
\def \rI {{\rm I}}

\def\pfour{{(p^2)^2}}

\begin{document}

\textwidth 170mm
\textheight 230mm
\topmargin -1cm
\oddsidemargin-0.8cm \evensidemargin -0.8cm
\topskip 9mm
\headsep9pt

\overfullrule=0pt
\parskip=2pt
\parindent=12pt
\headheight=0in \headsep=0in \topmargin=0in \oddsidemargin=0in

\vspace{ -3cm} \thispagestyle{empty} \vspace{-1cm}
\begin{flushright}
\end{flushright}
 \vspace{-1cm}
\begin{flushright} Imperial-TP-RR-03-2010 \\
PI-strings-200
\end{flushright}

\begin{center}

{\Large\bf
Quantum dispersion  relations for excitations\\
\vskip .2cm
 of long folded spinning superstring  in $AdS_5 \times S^5$
}

 \vspace{0.8cm} {
  S.~Giombi,$^{a,}$\footnote{sgiombi@perimeterinstitute.ca}
  R.~Ricci,$^{b,}$\footnote{r.ricci@imperial.ac.uk }
  R.~Roiban,$^{c,}$\footnote{radu@phys.psu.edu}
  and
  A.A.~Tseytlin$^{b,}$\footnote{Also at
   Lebedev  Institute, Moscow.\ \
   tseytlin@imperial.ac.uk }
 }\\
 \vskip  0.2cm

\small
{\em
$^{a}$
Perimeter Institute for Theoretical Physics,
Waterloo, Ontario, N2L 2Y5, Canada

 $^{b}$  The Blackett Laboratory, Imperial College,
London SW7 2AZ, U.K.

    $^{c}$ Department of Physics, The Pennsylvania State University,
University Park, PA 16802, USA
}

\normalsize
\end{center}

 \vskip 0.4cm

\begin{abstract}
We use $AdS_5 \times S^5$ superstring sigma model perturbation theory
to compute the leading  one-loop
corrections to the dispersion relations of the excitations near a long spinning string
 in AdS. This investigation is partially motivated by
the OPE-based approach to the computation of the expectation  value of null polygonal Wilson loops
suggested in arXiv:1006.2788.
Our results are in partial agreement with the recent
asymptotic Bethe ansatz  computation  in  arXiv:1010.5237.
In particular, we find that the heaviest AdS mode (absent in the ABA approach)
is stable  and has a corrected one-loop dispersion relation similar to
the other massive modes.
Its stability might hold also at the
next-to-leading order  as we  suggest using a unitarity-based argument.
\end{abstract}

\newpage

\section{Introduction \label{intro}}
\setcounter{footnote}{0}


The classical string solution describing a folded closed string
spinning around its center of mass  in $AdS_3 \subset   AdS_5 $ \ci{gkp,dev}
plays an important role in  many recent studies  of gauge--string duality.
In the large spin limit  when the string becomes very long and stretches
towards the boundary of $AdS_5$ the solution becomes very simple \ci{ft2,ftt}:
$t=\phi =  \k \tau, \ \ \r= \k \s,\  \ 0 < \s < {\pi \ov 2},
 \ \ \k \approx { 1 \ov \pi} \ln S $, i.e.
 homogeneous  or ``constant field strength'' one.\foot{Here the
$AdS_3$ metric in global coordinates is $ds^2 = - \cosh^2 \rho \  dt^2 + d \r^2 +
\sinh^2 \r \  d \phi^2$. This approximate  solution  is built out of 4
segments ($\r= \k \s,\  \ 0 < \s < {\pi \ov 2}$, etc.)
so that the string's length is $L= 2 \pi \k = 2 \ln S$. The ends of the string move
along null lines at the  boundary.}
   The
 resulting world surface    happens  to be    equivalent, via an analytic continuation and
 a global $SO(2,4)$ transformation,   to that of
  the null cusp solution  \ci{kru,amm},    explaining \ci{krtt,am2},  from a world-sheet
 perspective,
 the relation
 between the coefficient of the $\ln S$  term
  in the closed string  energy and the cusp
 anomaly.

The sum of  energies of small fluctuations around the solution  determines
the 1-loop correction to the folded string energy \ci{ft2}, but one can also study
the individual energies \ci{ft2,am2}  of these fluctuation modes
 which should represent, by analogy with the BMN case,
 the strong coupling limit of anomalous dimensions
 of the ``near-by'' gauge-theory operators.
 The dependence of these fluctuation mode energies on  the  gauge coupling $\l$ or
 string tension  came into focus recently \ci{ga,bas}
  in connection with the  investigation of the
 OPE for polygonal null Wilson loops \ci{ald}   related  to gluon scattering
 amplitudes in $\N=4$ SYM theory \ci{amm,alro}.
   In particular,
   ref. \ci{bas}  found   the expressions for
   the   dispersion relations for these modes 
   starting  from the   asymptotic Bethe  ansatz (ABA) of \ci{bes}.

Our  aim here will be to derive these dispersion relations to leading
non-trivial  order in  strong-coupling
$1 \ov \sql$   expansion  directly from the quantum \adss   superstring theory
and compare the results to the ABA predictions of \ci{bas}.\foot{Let us mention
 that the classical dispersion relation for  large  excitations
of folded string (analogous to giant magnons)
was constructed using classical integrability in \ci{dol}.}
Our approach  will be similar to that  used in a different context in
 \ci{zar}  and will be based on the AdS light-cone gauge \ci{mtt} formalism
  of our previous works \ci{us,uss}
 where we expanded near the null cusp open-string
 world surface.

Let us start with a summary of our results.
The direct  analysis  of  quadratic fluctuation operators
in the infinite spin  limit\foot{In general,  the  quadratic
 fluctuation operators near folded spinning string  have
 a  Lam\'e form and thus their spectra can be, in
principle,
explicitly determined also for finite spin $S$  \ci{bec}.
}
 reveals  \ci{ft2} the  presence
8+8 modes  with 2d relativistic dispersion relations
$E^2= p^2_1 + m^2$, where $p_1$ is the spatial component of the  momentum,
 $p_1 = {2 \pi n \ov L}= { n\ov \k}$:\foot{Our conventions here
are as follows:
$E= { 1 \ov \k}\er$, where  $\er$ is 2d energy,  so that  for a particle of mass
$s\,\k$ we have  $\er= \sqrt{ n^2   + s^2  \k^2}, \  \ E= \sqrt{ p^2_1 + s^2 } $.
Since for large  $S$ we have  $\k \gg 1$  we assume
that $n \gg 1$   so that the 2d spatial momentum $p_1$ is fixed in this  limit
(in this limit the spatial dimension of world sheet is decompactified).
}
one  bosonic mode with
 $m^2 = 4$ ($AdS_3$ fluctuation transverse to the string world sheet);
 2 bosonic  modes with $m^2 = 2$ ($AdS_5$ fluctuations transverse to $AdS_3$);
 5 bosonic modes with $m^2 = 0$  ($S^5$ fluctuations);
 8 fermionic  modes  with $m^2 = 1$.\foot{This interpretation of modes
 is found in conformal gauge. In general, it
  depends on  coordinate and gauge  choice.
For example, in conformal gauge there are two massless modes  in $AdS_3$
and 5 massless modes from $S^5$  two of which  are longitudinal and whose contribution
is  cancelled by that of
the conformal
ghosts. In the static gauge (with fluctuations of $t$ and $\r$ fixed)
or in the AdS light-cone gauge discussed below  all modes are physical  with
the massless modes coming  only from $S^5$.
}
As we find below,
the leading $1 \ov \sql$   corrections to the  dispersion relations of these
modes  may be written as  follows:
\be
\la{di}
&& E^2(p_1;\lambda) = \big[ p^2_1 + M^2(\l) \big]  \big[1 +  { \te{c \ov \sql}} p^2_1
 +   {\cal O}({\te{ 1\ov (\sql)^2 }})  \big] \ , \\
&&
M^2(\l) = m^2 + {\te { q \ov \sql}} +   {\cal O}({\te{ 1\ov (\sql)^2 }}) \ ,
\la{mc} \\
&&
m^2_{_{ AdS_3}} = 4,\ \  m^2_{_{\perp  AdS_3}} = 2,\ \
 m^2_{_{S^5}} = 0,\ \   m^2_{_{F}} =1  \ , \la{v}  \\
 &&
 q_{_{ AdS_3}} = 0 ,\ \  \ q_{_{\perp AdS_3}} =  - \pi \ , \ \ \
   q_{_{S^5}} = q_{_{F}} =0  \ , \la{vv} \\
&&   c_{_{   AdS_3}} = - \ha \pi ,\ \  c_{_{\perp AdS_3}} =      - \pi \ , \ \ \
c_{_{S^5}} = - \frac{7}{3} \ , \ \ \ \  c_{_{F}} =- 2 \pi  \ , \la{vvv}
\ee
We observe that the masses (defined as  the values of energy at vanishing momentum)
of the $AdS_3$ mode, the fermions  and the $S^5$ scalars
are not renormalized:
$E_{_{ AdS_3}} (0;\lambda)= 2, \ \ E_{_{ F}} (0;\lambda)=1, \ \  E_{_{ S^5}} (0;\lambda)=0$.
Also, while for the  2  transverse $AdS_5$ modes the mass gets renormalized,
we find that $E_{_{\perp  AdS_3}} (\pm {\rm i};\lambda )= 1$  (to leading  order in ${1 \ov \sql}$
we consider).
These exact relations  are  in agreement with the expectations \ci{am2}
based  on the interpretation of these modes as Goldstone particles associated with
partial breaking of the original global $SO(2,4) \times SO(6)$ symmetry by the classical
solution.
 According to \cite{bas}, the ABA predicts that the relations $E_{_{\perp  AdS_3}}
 (\pm {\rm i};\lambda )= 1$, $E_{_{ F}} (0;\lambda)=1$ 
 should hold for any  $\lambda$  and 
  $E_{_{ F}}
  (\pm {\rm i};\lambda)=0$ should be valid to all orders in 
  strong coupling expansion.

The leading mass shift of the two ``transverse'' $AdS_5$ modes
($M^2_{\perp AdS_3}=2 - { \pi \ov \sql} +  {\cal O}(\te{ 1\ov (\sql)^2 })$)
is in exact agreement with the ABA  result of \ci{bas}.
The  results for the coefficients
$c_{_{\perp AdS_3}} =   - \pi= \ha c_{_{F}}$ also agree with  those of \ci{bas}.

This matching is not of course  totally unexpected as the ABA  is already
 known to be in
 agreement with the
semiclassical string theory,\foot{The one- and two-loop corrections to the cusp anomaly
match; the leading-order energies of fluctuations near folded string
are  also   captured by
the semiclassical ABA  or through  algebraic curve
 considerations  (see, {\it e.g.}, \ci{grom}).}
 though subleading
 corrections to dispersion relations test  more than just  the  ``1-loop''
  or ``quadratic'' fluctuation Lagrangian but also the
  interaction vertices and thus are
  similar in spirit to the 2-loop string corrections to the ``ground-state'' energy.

  At the same time, we also find   some differences  compared to the results of  \ci{bas}.
  First, at very low energies 
  the massless $S^5$ modes   
  are expected  to decouple from the massive   modes and thus be
  described  \ci{am2}  by 
  an effective 
  $O(6)$ sigma model  whose  asymptotic states are  6  massive scalars with 
   non-perturbatively generated  mass
  $M^2_{_{S^5}} \sim m^2 e^{- \frac{1}{2} { \sql }}$ 
  ($m$ is the mass scale of our massive modes  or an effective UV cutoff).
  Our perturbative  computation of the 2-point function does not, of course, 
  capture this effect.
The asymptotic $O(6)$ states considered in \ci{bas}
 are thus different from  our 5  massless scalars and so 
one may think  that  there is no a priori reason why
 the value of $c_{_{S^5}} = - 8 \pi c_{\rm scalar} = { 1 \ov  12
 \pi (12)^{1/4} } [ \Gamma( \frac{1}{4})]^4
 \approx  2.46$  found in \ci{bas}  should match our value
  in \rf{vvv}.
 Still, we do not understand the origin of this disagreement
  assuming we are  considering the same  range of momenta and
coupling values.\foot{
  At large $\l$ non-perturbative effects should  not be relevant
  so unless one considers special  scaling of $p_1$
  one would expect to find matching.}

  

More importantly, the heaviest $AdS_3$ mode is apparently absent in the
all-order ABA analysis  of \ci{bas}. One possible reason for  why it may have been  missed
 is that
it should  be identified  not with  twist one   but with a  higher twist
excitation \ci{bas}.
Another   possible explanation  suggested in \ci{ald}  is that this mode may decay
into a pair of fermions (notice that $m_{_{AdS_3}}= 2 m_{_F}=2$)
 and thus  may disappear from the spectrum at finite $\l$
by analogy with a similar  proposal \ci{zar} for the heavier
 BMN-type excitation in $AdS_4 \times \mathbb {CP}^3$ context. In \ci{zar}
 it was argued that  the corresponding
loop-corrected propagator  has no longer  a pole  but  a branch cut
 and therefore does  not describe an asymptotic state.

We do not  find, however,  evidence in support of  this  scenario in the present case:
the corrected  propagator still has a real pole  described by \rf{di}.
One may also give a unitarity-based  argument that the decay
of the heavy $AdS_3$ mode into  two
fermions will not happen   at 1-loop order.
    One may   wonder  though
  if this  heavy mode may  still be interpreted as a stable threshold bound state
 of two  fermions  and thus should  not be considered as a separate excitation
 in the ABA spectrum.
A hint in that direction is  the observation
 that the  corrected  dispersion  relations   \rf{di}--\rf{vvv}  for
  the two  modes are related
 in the same way as their tree-level  counterparts:
 \be \la{mff}
 E_{_{ AdS_3}} (p_1; \l )= 2  E_{_{ F}} (\ha p_1; \l ) \  .
 \ee
 It remains to be seen if that ``bound state'' interpretation can be
 given some precise sense.

It is possible  that the fate of this heaviest bosonic mode  can be clarified
by starting with a folded  spinning string solution with an extra
orbital momentum $J=\sql \nu $ in $S^5$ \ci{ft2}.  In that case the fluctuation spectrum
 contains the following  physical modes in conformal gauge \ci{ftt}:\foot{In
 the static
 gauge ($\tilde t = \tilde \rho=0$)
   where the longitudinal massless modes are already
  gauge fixed  the two modes with energies
 $E^{(\pm)}_{_{AdS_3}}$  originate from mixing
 of the angular fluctuations in the $AdS_3$ and in the $S^5$ (those in the angular momentum
  carrying  directions). Similar picture is found in the AdS light-cone
  gauge \ci{us,uss}.}
  \be \la{enn}
&&E^{(\pm)}_{_{AdS_3}}
 =\sqrt{p^2_1 +  2   \pm   2 \sqrt{ 1 + { \te{ \ell^2 \ov \ell^2 + 1} }  p^2_1 }}  \ , \ \
\ \ \ \ \  E^{\times 2}_{_{\perp AdS_3}} = \sqrt{p^2_1 +  2  - {\te { \ell^2 \ov \ell^2 + 1} }
  } \ , \\  &&
E^{\times 4} _{_{S^5}}   = \sqrt{p^2_1 + {\te{  \ell^2 \ov \ell^2 + 1} } } \ ,
\ \ \ \ \ \ \ \ \  \ \ \ \   \ \ \ \ \ \
E^{\times 8} _{_{F}} = \sqrt{p^2_1 +  1 } \ ,\la{ennn}  \ee
where $p_1 = {2 \pi n \ov L}= { n \ov \k},  \ \
L= 2\pi  \k = 2 \pi \sqrt{\J^2 + { 1 \ov \pi^2} \ln^2 S } =
 2 \ln S  \sqrt{1 + \ell^2},
 \ \ \k^2 = \m^2 + \nu^2 ,\ \ \mu = {1 \ov \pi} \ln S , \ \
\ell \equiv  { \nu \ov \m} $.
Here the unphysical modes are
 one time-like massless mode in $AdS_5$ and one space-like  massless mode in $S^5$.
The $AdS_3$ mode with energy $E^{(-)}_{_{ AdS_3}}$
becomes a massless field with relativistic dispersion relation in the limit $J \to 0$, i.e. $\ell \to 0$.

The heaviest  mode with energy   $E^{(+)}_{_{ AdS_3}}$  may  then be
interpolated to a large $J$  state in the  sl(2) sector
that is visible  also at weak coupling
 while
 for small $J$  it reduces back to the above stable $m=2$ mode.
Modulo subtleties of the limits involved, that suggests
that this mode should be there in the ABA description at all $\l$.
\foot{Note, however,  that in the  small $\ell$ expansion
the tree-level mass  of the heavier mode
($E^{(+)}_{_{ AdS_3}}= \sqrt{p_1^2 + 4} \big[ 1 + { p_1^2 \ov 2 (p_1^2
+ 4)}\ell^2 + ... \big]$)
is still $2$, i.e. is still  equal to  the
sum of the two fermion masses, implying again   a possibility of a
threshold decay.}

The structure of the rest of this paper is as follows.
In section 2 we review the \adss superstring action in the AdS light-cone
gauge and expand it near a classical solution representing the null cusp surface which is equivalent
to the  large spin limit of the folded string.
In section 3  we compute the 1-loop corrections  to the 2-point functions of the
fluctuation fields in the action. This allows us to extract corrections \rf{di}--\rf{vvv}
to their
dispersion relations.
In section 4  we shall comment  on
stability of the heaviest $m=2$
mode  pointing out the difference between the  present case
and  the one discussed in \ci{zar}.
Some concluding remarks  are in  section 5.

\section{Review of the AdS light-cone gauge action}

In this section we briefly review the structure of the $AdS_5\times S^5$
superstring action in the AdS light-cone gauge \cite{mtt}, in particular
 its expansion around the null cusp solution \cite{us}.

The AdS light-cone gauge is adapted to the  Poincar\'e patch of \adss
($m=0,1,2,3; \ M= 1,...,6$)
\bea
&&ds^2 = z^{-2} ( dx^m dx_m   + dz^M dz^M)
= z^{-2} ( dx^m dx_m   + dz^2) + du^M du^M \ ,\la{m} \\
&&   x^m x_m =  x^+ x^-  + x^* x   \ , \ \ \ \    \
x^\pm = x^3 \pm x^0 \ , \ \ \ \ \ \  x= x^1 + \mathrm{i} x^2 \ ,
\la{x} \\
&& z^M = z u^M \ , \ \ \ \ u^M u^M= 1 \ , \ \ \ \ \ \
z= (z^M z^M)^{1/2} \equiv  e^\phi   \ . \la{z}
\eea
Starting with the superstring action of \cite{mt} in these coordinates, the
 AdS light-cone gauge is defined by fixing $\k$-symmetry by the condition
 $ \Gamma^+ \theta^I=0$
on the two 10-d  Majorana-Weyl  fermions as well as imposing
\begin{equation}
  \label{ga}
  \sqrt{-g} g^{\alpha\beta} ={\rm diag}(-z^2, z^{-2})\ , \qquad \qquad x^+ = p^+ \tau \ ,
\end{equation}
where   $g_{\alpha\beta}$ is the 2-d metric.
The latter conditions fix completely the 2d diffeomorphism invariance and decouple the unphysical modes from the theory.

We will work with the Euclidean version of the worldsheet action, which may be obtained by redefining $\tau \to  -\mathrm{i} \tau, \ p^+ \to  \mathrm{i} p^+$. After setting $p^+=1$, the gauge fixed $AdS_5\times S^5$ superstring action in Euclidean signature takes the form
\be
\label{s}
I &=&  \frac{1}{2} T \int d \tau \int
 d \sigma \; \mathcal{L}_E\ , \quad  \quad  \quad T =
\frac{R^2}{2 \pi \alpha'} = \frac {\sqrt{\lambda}}{2 \pi} \ , \\
\mathcal{L}_E& =& \dot{x}^* \dot{x} + (\dot z^M  + \mathrm{i}  z^{-2} z_N
\eta_i {\rho^{MN}}^i{}_j \eta^j)^2  + \mathrm{i}  (\theta^i \dot{\theta}_i +
       \eta^i\dot{\eta}_i - h.c.) -  z^{-2} (\eta^2)^2 \nonumber \\
   &&+  z^{-4} ( x'^*x'  + {z'}^M {z'}^M) + 2 \mathrm{i} \Big[\
       z^{-3}\eta^i \rho_{ij}^M z^M (\theta'^j - \mathrm{i}
       z^{-1} \eta^j  x') + h.c.\Big]\;.  \label{euc}
       \ee
The fermions are
complex $\th^i = (\th_i)^\dagger,$ $\eta^i = (\eta_i)^\dagger\ $
$(i=1,2,3,4)$ transforming in the fundamental representation of $SU(4)$.
The $\rho^{M}_{ij} $ are  off-diagonal blocks of
six-dimensional gamma matrices in chiral representation and
$(\rho^{MN})_i^{\hphantom{i} j} = (\rho^{[M}
  \rho^{\dagger N]})_i^{\hphantom{i} j}$ and
$(\rho^{MN})^i_{\hphantom{i} j} = ( \rho^{\dagger [M}
  \rho^{N]})^i_{\hphantom{i} j}$ are  the $SO(6)$ generators.

The Euclidean action admits the following classical solution
\bea
{z}=\sqrt{\frac{\tau}{\sigma}}\ , \qquad x^+ =  \tau\ , \qquad
 x^-  = - { 1 \ov 2\sigma} \ , \ \ \ \ \ \ \ x_1=x_2=0 \,.
\label{cusp}
\eea
This is nothing but the null cusp background \ci{kru,krtt} written in this
light-cone gauge. It describes a euclidean  world surface of an open string
ending on the AdS boundary (we assume that $\tau$ and $\sigma$
 change from 0 to $\infty$). Since $x^+x^-=0$ at $z=0$ this surface ends on a
 null cusp. As was already mentioned above, this light-like cusp solution is
 related  by an analytic continuation
and a global conformal transformation to the  infinite spin limit
of the  folded string solution.

Fluctuations of fields around this classical solution can be defined by \cite{us}
\bea\la{flu}
&& z=\sqrt{\frac{\tau}{\sigma}}\ {\tilde z} \ , \ \ \ \ \ \ \ \
{\tilde z} = e^{\tilde \phi}= 1 + \tilde \phi  +\dots~,\ \ \
 z^M=\sqrt{\frac{\tau}{\sigma}}\ {\tilde z}^M \ , \ \ \ \
{\tilde z}^M = e^{\tilde \phi} \tilde u^M  \\
&&
{\tilde u}{}^{a}=  \frac{y^{a}}{1+\frac{1}{4}y^2}~, \ \ \ \
{\tilde u}{}^{6} =  \frac{1-\frac{1}{4}y^2}{1+\frac{1}{4}y^2}  \ , \ \ \ \ \ \ \ \ \
~~~~ y^2\equiv \sum_{a=1}^5 (y^a)^2\ , \ \ \ \ \ a=1,...,5 \ , \la{u} \\
&&
x = \sqrt{\frac{\tau}{\sigma}} \ {\tilde x}
~,~~~~~~
\theta=\frac{1}{\sqrt{2\sigma}}{\tilde\theta}
~,~~~~~~
\eta=\frac{1}{\sqrt{2\sigma}}{\tilde\eta}\ . \la{xx}
\eea
A further redefinition of the worldsheet coordinates $(\tau,\sigma) \to (t,s)$
which makes the induced world-sheet metric conformally flat
\footnote{Notice that compared to \cite{us} the $(t,s)$ coordinates are defined here with an extra factor of 2, so that the masses of the fluctuations fields are $m^2=(4,2,1)$ instead of the values $m^2=(1,1/2,1/4)$ used in \cite{us}.}
\be\la{oo}
t=2\ln \tau~,~~~~~~s=2\ln \sigma~,~~~~~~~~~~~dt ds=4\frac{d\tau d\sigma}{\tau\sigma}
~,~~~~~~~2\tau\partial_\tau=\partial_t
~,~~~~~~~~2\sigma\partial_\sigma=\partial_s\ ,
\ee
leads then to the following Euclidean action:
\be\label{action}
S &=&\frac{\sqrt{\lambda}}{4\pi}\int dt \int^\infty_{-\infty} ds\ {\cal L}\ ,\\
\label{Lagrangian}
{\cal L}  &=&
\big|\partial_t {\tilde x}+{\tilde x}\big|^2+
\frac{1}{{\tilde z}^{4}} \big|\partial_s {\tilde x} -{\tilde x}\big|^2
\\
&+& \big( \partial_t {\tilde z}^M +{\tilde z}^M  +
\frac{{\rm i} }{{\tilde z}^2}
{\tilde \eta}_i  (\rho^{MN}){}^i{}_j {\tilde \eta}^j  {\tilde z}_N \big)^2
+ \  \frac{1}{{\tilde z}^{4}} \big(\partial_s{\tilde z}^M -{\tilde z}^M \big)^2
 \cr
&+&  {\rm i}
({\tilde \theta}^i \partial_t{\tilde \theta}_i
+{\tilde \eta}^i\partial_t{\tilde \eta}_i + {\tilde \theta}_i
\partial_t{\tilde \theta}^i
+{\tilde \eta}_i\partial_t{\tilde \eta}^i)
-  \ \frac{1}{{\tilde z}^{2}} ({\tilde \eta}^2)^2
\cr
&+&  2{\rm i}\ \Bigl[\ \frac{1}{{\tilde z}^{3}}{\tilde \eta}^i (\rho^M)_{ij} {\tilde z}^M
(\partial_s{\tilde \theta}^j -{\tilde \theta}^j
- \frac{{\rm i}}{{\tilde z}} {\tilde \eta}^j
 (\partial_s x-x))\cr
&&~~~~~~~~~~~~~~~~~~~~~
+\frac{1}{{\tilde z}^{3}}{\tilde \eta}_i (\rho^\dagger_M)^{ij} {\tilde z}^M
(\partial_s {\tilde \theta}_j - {\tilde \theta}_j
+ \frac{{\rm i}}{{\tilde z}} {\tilde \eta}_j(\partial_sx^* -x^*))\Bigr] \ .
\nonumber
\ee
The spectrum of excitations can be read off from the quadratic part of the
fluctuation Lagrangian:
\be
{\cal L}_2  &=&\partial_{\alpha}\tilde\phi\partial_{\alpha}\tilde\phi +4{\tilde \phi}^2 +
\partial_{\alpha} {\tilde x}\partial_{\alpha} {\tilde x}^*
+2{\tilde x}{\tilde x}^{*}
+\partial_{\alpha}y^a\partial_{\alpha}y^a\no
 \\
&+&  2{\rm i}\;
({\tilde \theta}^i \partial_t{\tilde \theta}_i
+{\tilde \eta}^i\partial_t{\tilde \eta}_i
)
+2{\rm i}\; {\tilde \eta}^i (\rho^6)_{ij}
        (\partial_s{\tilde \theta}^j -{\tilde \theta}^j)
+
2{\rm i}\; {\tilde \eta}_i (\rho^\dagger_6)^{ij}
(\partial_s {\tilde \theta}_j -{\tilde \theta}_j) ~,
\la{qua}
\ee
where $\partial_\alpha=(\partial_0, \partial_1) = (\partial_t, \partial_s)$.
Thus we see that the bosonic modes are:
(i) one field (${\tilde \phi}$) with $m^2=4$; (ii)  two
 fields ($\tilde x,\tilde x^*$) with $m^2=2 $; (iii)  five
fields ($y^a$) with $m^2= 0$. Notice that unlike the  case of the
 conformal gauge, here the bosonic propagator is diagonal. The tree level 2-point functions
 in momentum space are simply ($p^2 \equiv p_\alpha p_\alpha= p_0^2 + p_1^2$,
  \  $p_0=p_t, \
 p_1=p_s$)
\begin{equation}
\langle \tilde{x}(p)\tilde{x}^*(-p)\rangle_0 = \frac{2\pi}{\sqrt{\lambda}}\frac{2}{p^2+2}\,,\qquad
\langle \tilde{\phi}(p) \tilde{\phi}(-p) \rangle_0=
\frac{2\pi}{\sqrt{\lambda}}\frac{1}{p^2+4}\ , \qquad
\langle y^a(p) y^b(-p) \rangle_0 = \frac{2\pi}{\sqrt{\lambda}}\frac{\delta^{ab}}{p^2}\,.
\end{equation}
In fact, one can see that the boson 2-point function matrix should remain diagonal to any loop order, because the Lagrangian has a global $SO(2)_{\tilde{x}}\times SO(5)_y$ symmetry which prevents mixing between the bosonic fields.

Computing the determinant of the fermionic kinetic operator in momentum space \cite{us}, one finds that the 8 physical fermionic degrees of freedom all have $m^2=1$, as required by the $SO(6)$ symmetry of the null cusp background \cite{am2}. The non-trivial tree level 2-point functions read
\begin{eqnarray}
&&\langle \theta^i(p)\eta^j(-p) \rangle_0 =
 -\frac{2\pi}{\sqrt{\lambda}}\frac{p_1-\mathrm{i} }{p^2+1}(\rho_6^{\dagger})^{ij}\,,
\ \ \ \ \qquad
\langle \theta_i(p)\eta_j(-p) \rangle_0 = -\frac{2\pi}{\sqrt{\lambda}}\frac{p_1-\mathrm{i}
 }{p^2+1}\rho^6_{ij}\ , \cr
&&\langle \theta^i(p)\theta_j(-p) \rangle_0 =\langle \eta^i(p)\eta_j(-p) \rangle_0=-\frac{2\pi}{\sqrt{\lambda}}\frac{p_0}{p^2+1}\delta^i_j\,.
\label{fermions-tree}
\end{eqnarray}
To compute 1-loop corrections to the  2-point functions
 we need to expand the above fluctuation action to quartic order.
 The cubic and quartic interaction vertices can be read off from
\bea
&&{\cal L}_3=
-4\tilde\phi\, |\partial_s\tilde x-\tilde x|^2+2\tilde\phi\ [(\partial_t\tilde \phi)^2-(\partial_s\tilde \phi)^2]+2\tilde\phi\ [(\partial_t y^a)^2-(\partial_s y^a)^2]\cr
&&\hspace{9truemm}
-4 \mathrm{i}\tilde\phi\ [\tilde\eta^i\,(\rho^6)_{ij}
(\partial_s\tilde\theta^j-\tilde\theta^j)+\tilde\eta_i\,(\rho_6^{\dagger})^{ij}
(\partial_s\tilde\theta_j-\tilde\theta_j)]\cr
&&\hspace{9truemm}
+2\mathrm{i}y^a\ [\tilde\eta^i\,(\rho^a)_{ij}
(\partial_s\tilde\theta^j-\tilde\theta^j)+\tilde\eta_i\,(\rho_a^{\dagger})^{ij}
(\partial_s\tilde\theta_j-\tilde\theta_j)]
+2\mathrm{i}\tilde\eta_i\,(\rho^{a6}){}^i{}_j\tilde\eta^j\partial_t y^a\cr
&&\hspace{9truemm}
+2\tilde\eta^i\,(\rho^6)_{ij}\tilde\eta^j(\partial_s \tilde x-\tilde  x)
-2\tilde\eta_i\,(\rho_6^{\dagger})^{ij}\tilde\eta_j(\partial_s \tilde x^*-\tilde  x^*)
\label{cubic}
\eea
and
\bea
&&{\cal L}_4=
+8\,\tilde\phi^2\, |\partial_s \tilde  x-\tilde x|^2+2\tilde\phi^2\
[\partial_{\alpha}\tilde\phi\partial_{\alpha}\tilde\phi
+\frac{2}{3}\tilde\phi^2]
+2\tilde\phi^2\ \partial_{\alpha} y^a\partial_{\alpha} y^a
-\frac{1}{2}y^a y^a\, \partial_{\alpha} y^b \partial_{\alpha} y^b\cr
&&\hspace{9truemm}
+\mathrm{i}(4\tilde\phi^2-\, y^a y^a)\,[\tilde\eta^i\,(\rho^6)_{ij}
(\partial_s\tilde\theta^j-\tilde\theta^j)+\tilde\eta_i\,(\rho_6^{\dagger})^{ij}
(\partial_s\tilde\theta_j-\tilde\theta_j)]\cr
&&\hspace{9truemm}
-4\mathrm{i}\tilde\phi\,y^a\ [\tilde\eta^i\,(\rho^a)_{ij}
(\partial_s\tilde\theta^j-\tilde\theta^j)+\tilde\eta_i\,(\rho_a^{\dagger})^{ij}
(\partial_s\tilde\theta_j-\tilde\theta_j)]\cr
&&\hspace{9truemm}
-6\tilde\phi\,[\tilde\eta^i\,(\rho^6)_{ij}\tilde\eta^j(\partial_s \tilde x-\tilde  x)
-\tilde\eta_i\,(\rho_6^{\dagger})^{ij}\tilde\eta_j(\partial_s \tilde x^*-\tilde  x^*)]\cr
&&\hspace{9truemm}
-2\mathrm{i}y^a\partial_t y^b\,\tilde\eta_i\,(\rho^{ab}){}^i{}_j\tilde\eta^j
-\tilde\eta_i\,(\rho^{a6}){}^i{}_j\tilde\eta^j\,\tilde\eta_k\,(\rho^{a6}){}^k{}_l\tilde\eta^l
-\tilde\eta^i\,\tilde\eta_i\,\tilde\eta^j\,\tilde\eta_j\,.
\label{quartic}
\eea

\section{The 2-point functions at 1-loop order}

\begin{figure}
\begin{center}
\includegraphics[width=115mm]{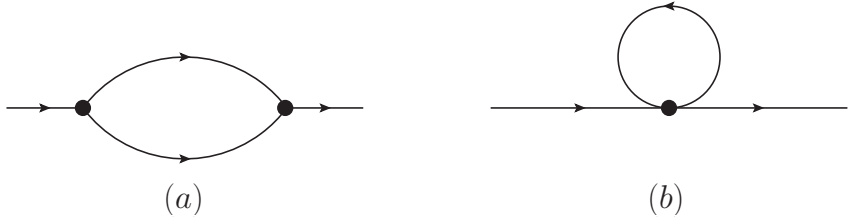}
\parbox{13cm}{\caption{The  topologies  of 1PI graphs
contributing to the two point function at 1-loop.}
\label{1PI}}
\end{center}
\end{figure}

\begin{figure}
\begin{center}
\includegraphics[width=55mm]{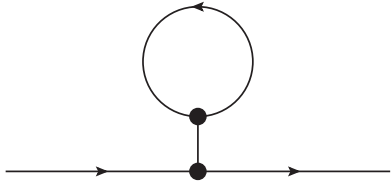}
\parbox{13cm}{\caption{The tadpole topology of the graph contributing
to the two point
function at 1-loop. The bubble is fermionic and it is connected to the
propagator
by a zero-momentum $\tilde\phi$ line.}
\label{non-1PI}}
\end{center}
\end{figure}

In this section we present the details of our computation of the 1-loop correction
 to the 2-point functions of the fluctuation fields. The calculation is, in principle,
  straightforward: after reading off the Feynman rules from (\ref{cubic}),(\ref{quartic}) we shall simply compute the relevant 1-loop self-energy diagrams. A peculiarity of the light-cone action expanded around the null cusp, which was noticed in \cite{us}, is that the fluctuation field $\tilde{\phi}$ acquires a non-trivial UV divergent one-point function at 1-loop
\bea
\langle \tilde\phi \rangle = -\frac{8\pi}{\sqrt{\lambda}}
\int \frac{d^2q}{(2\pi)^2}\frac{q_1^2+1}{q^2+1}
\eea
which is due to the $\tilde\phi\tilde\eta\tilde\theta$-interaction in (\ref{cubic}). The
presence of this tadpole implies that, besides the one-particle irreducible diagrams of
Fig.~\ref{1PI}, we should include the one-particle reducible topology in Fig.~\ref{non-1PI}.
Similarly to the calculation of the partition function \cite{us}, the inclusion of this diagram is
important for the cancellation of divergences.
Below we shall use the following notation for the  1-loop integrals:
\be \rI[m^2] = \int \frac{d^2q}{(2\pi)^2} \frac{1}{q^2+m^2} \ , \la{dee}\ee
and
\begin{eqnarray}
G[m_1^2,m_2^2;p^2]=\int \frac{d^2q}{(2\pi)^2} \frac{1}{(q^2+m_1^2)((p-q)^2+m_2^2)}
=\frac{1}{4\pi} \int_0^1 dx \frac{1}{p^2 x(1-x)+m_1^2 x +m_2^2 (1-x)}\,.
\label{GG}
\end{eqnarray}
The latter is UV finite and equal to
\be \la{mmm}
G[m_1^2,m_2^2;p^2]=
\frac{\ln\ \frac{p^2+m_1^2+m_2^2+\sqrt{(p^2+m_1^2+m_2^2)^2-4m_1^2m_2^2}}
{p^2+m_1^2+m_2^2-\sqrt{(p^2+m_1^2+m_2^2)^2-4m_1^2m_2^2}}}{4\pi\sqrt{(p^2+m_1^2+m_2^2)^2-4m_1^2m_2^2}}\,.
\ee
In particular, we find  that ($|p| = \sqrt{ p^2}$)
\be
G[m^2,m^2;p^2]=
  \frac{{\rm arcsinh}\frac{|p|}{2m}}{\pi \ |p|\sqrt{p^2 + 4 m^2}}\, .
\ee

\subsection{The $m=\sqrt{2}$ bosons}

Let us start from the correction to the $\tilde{x}\tilde{x}^{*}$ 2-point function,
which is the one involving the smaller number of diagrams due to the simplicity
of the relevant vertices. The purely bosonic contribution to the 1-loop correction is
\begin{eqnarray}
\langle \tilde{x}(p)\tilde{x}^*(-p)\rangle_1^{\rm bose}=
\left(\frac{2\pi}{\sqrt{\lambda}}\right)^2\,32 \frac{p_1^2+1}{(p^2+2)^2}
\left[\int \frac{d^2q}{(2\pi)^2} \frac{(p_1-q_1)^2+1}{(q^2+4)((p-q)^2+2)}
-\frac{1}{2}\int \frac{d^2q}{(2\pi)^2}\frac{1}{q^2+4}\right]\,.
\label{xx-B}
\end{eqnarray}
The first term here arises from the 3-vertex $\tilde{x}\tilde{x}^*\tilde{\phi}$, while
the second one from the 4-vertex $\tilde{x}\tilde{x}^*\tilde{\phi}^2$.

The contributions involving fermions come from the 3-vertex
$\tilde{x}\tilde\eta_i\tilde\eta^j$ and from the diagram in Fig.~\ref{non-1PI}
obtained by attaching the $\tilde{\phi}$ tadpole to the propagator using the
$\tilde{x}\tilde{x}^*\tilde{\phi}$ vertex. The sum of the two diagrams read
\begin{eqnarray}
\langle \tilde{x}(p)\tilde{x}^*(-p)\rangle_1^{\rm fermi}=
\left(\frac{2\pi}{\sqrt{\lambda}}\right)^2\,32 \frac{p_1^2+1}{(p^2+2)^2}
\left[\int \frac{d^2q}{(2\pi)^2} \frac{(q_0-p_0)q_0}{(q^2+1)((p-q)^2+1)}
-\int \frac{d^2q}{(2\pi)^2}\frac{q_1^2+1}{q^2+1}\right]\,.
\label{xx-F}
\end{eqnarray}
One can see that the purely bosonic part eq.~(\ref{xx-B}) and the fermionic part
eq.~(\ref{xx-F}) are separately free of logarithmic UV divergences. In (\ref{xx-F}),
the tadpole contribution is important to achieve this.

The 1-loop integrals above may be evaluated in several different ways. We will
choose to first perform a tensor reduction of the numerator factors,
as described,  for example,  in the Appendix of \cite{us}. We find
\be
&&\langle \tilde{x}(p)\tilde{x}^*(-p)\rangle_1^{\rm bose}=
\left(\frac{2\pi}{\sqrt{\lambda}}\right)^2\, \frac{p_1^2+1}{(p^2+2)^2}
\Bigg{[}
\frac{8}{\pfour}\left(\left(p^2+2\right)^2
+4p^2\right)(p_1^2-p_0^2)G[2, 4; p^2]\cr
&&~~~~~~~~~~~~~~~~~~~~~~~~~~~~~~~~~~~
-\frac{8}{\pfour}\left(p^2+2\right)(p_0^2-p_1^2)(\rI[4]-\rI[2])
\Bigg{]}
\ee
for the bosonic contribution and
\be
\langle \tilde{x}(p)\tilde{x}^*(-p)\rangle_1^{\rm fermi}=
\left(\frac{2\pi}{\sqrt{\lambda}}\right)^2\,\frac{p_1^2+1}{(p^2+2)^2}
\Bigg{[}-\frac{8}{p^2}(\pfour+4p_1^2)G[1,1;p^2]\Bigg{]}
\ee
for the 
fermionic contribution.

Putting the two pieces together, the 1-loop correction to the $\tilde{x}\tilde{x}^*$
2-point function is
\begin{eqnarray}
\langle \tilde{x}(p)\tilde{x}^{*}(-p)\rangle_1 &=& \left(\frac{2\pi}{\sqrt{\lambda}}\right)^2
\frac{p_1^2+1}{(p^2+2)^2} \Bigg[
\frac{8}{\pfour}\left(\left(p^2+2\right)^2
+4p^2\right)(p_1^2-p_0^2)G[2, 4; p^2]\cr
&&
-\frac{8}{p^2}
\left(\pfour+4 p_1^2\right)G[1,1;p^2]
-\frac{8}{\pfour}\left(p^2+2\right)(p_0^2-p_1^2)(\rI[4]-\rI[2])
\Bigg]
\cr
&\equiv& \left(\frac{2\pi}{\sqrt{\lambda}}\right)^2 \frac{2}{(p^2+2)^2}\,F^{(1)}_{\tilde{x}
\tilde{x}^*}(p_0,p_1) \la{xexx}
\end{eqnarray}
Thus  the resummed 2-point function in the 1-loop approximation is
\begin{equation}
\langle \tilde{x}(p)\tilde{x}^{*}(-p)\rangle = \frac{2\pi}{\sqrt{\lambda}}\ \frac{2}{p^2+2
-\frac{2\pi}{\sqrt{\lambda}}F^{(1)}_{\tilde{x}\tilde{x}^*}(p_0,p_1)}\,.
\end{equation}
The pole of the propagator is at $p^2=-2+{\cal O}(\frac{1}{\sqrt{\lambda}})$;
therefore,  to the 1-loop order we only need the ``on-shell" value of
$F^{(1)}_{\tilde{x}\tilde{x}^*}(p_0,p_1)$ which, using \rf{mmm}, i.e.
 $G[1,1;p^2=-2]=4G[2,4;p^2=-2]=\frac{1}{8}$, is simply
\begin{equation}
F^{(1)}_{\tilde{x}\tilde{x}^*}(p_0,p_1)|_{p^2=-2} = \frac{1}{2}\left(p_1^2+1\right)^2\,.
\label{Fx}
\end{equation}
The 1-loop corrected dispersion relation for the 2 bosons of mass $m^2=2$ is then
\begin{equation}
p^2=-2+\frac{\pi}{\sqrt{\lambda}}\left(p_1^2+1\right)^2+{\cal O}(\frac{1}{\lambda})\,.
\end{equation}
Wick-rotating to the Minkowski signature $(p_0,p_1)=(i E,p_1)$, this gives the dispersion relation
\begin{eqnarray}\la{tamm}
&&E^2=p_1^2+2-\frac{\pi}{\sqrt{\lambda}}(p_1^2+1)^2+{\cal O}(\frac{1}{\lambda})\cr
&&=\left(p_1^2+2-\frac{\pi}{\sqrt{\lambda}}\right)\left[1-\frac{\pi}{\sqrt{\lambda}}p_1^2
\right]+{\cal O}(\frac{1}{\lambda})\,.
\end{eqnarray}
This precisely agrees with the result of \cite{bas}.

\subsection{The $m=2$ boson \label{highestmass}}

Let us now consider the correction to the $\langle
\tilde{\phi} \tilde{\phi} \rangle$ 2-point function.
This calculation is especially interesting because this mode is absent in the analysis of \cite{bas}.

Computing the relevant diagrams, one finds that the
contributions corresponding to the topology involving two 3-vertices, Fig.~\ref{1PI}$(a)$, are
 (below $s^{\alpha\beta}=$diag$(1,-1)$)
\begin{eqnarray}
&&\langle\tilde\phi(p)\tilde\phi(-p)\rangle_1^{V_3V_3}=\left(\frac{2\pi}{\sqrt{\lambda}}\right)^2 \frac{1}{(p^2+4)^2}\int \frac{d^2q}{(2\pi)^2}\Bigg{[} s^{\alpha\beta}s^{\gamma\delta}\frac{2 q_{\alpha} q_{\gamma} (q_{\beta} q_{\delta}+2 p_{\beta} q_{\delta}-3p_{\beta} p_{\delta})+4 q_{\alpha} p_{\beta} p_{\gamma} p_{\delta}}{(q^2+4)((p-q)^2+4)}\cr
&&~~~~~~~~~~~~~~~~~~~~~~~~~~~~~~~+16\frac{(q_1^2+1)((p_1-q_1)^2+1)}{(q^2+2)((p-q)^2+2)}
+10 s^{\alpha\beta}s^{\gamma\delta} \frac{q_{\alpha} q_{\gamma} (p_{\beta}-q_{\beta})(p_{\delta}-q_{\delta})}{q^2 (p-q)^2}\cr
&&~~~~~~~~~~~~~~~~~~~~~~~~~~~~~~~-32 \frac{((p_1-q_1)^2+1)(q_1^2+1+q_0 (p_0-q_0))}{(q^2+1)((p-q)^2+1)}\Bigg{]}\,.
\label{phi-V3V3}
\end{eqnarray}
Here the first two lines include all purely bosonic contributions, while the last line
corresponds to the diagram with a fermion loop. The contributions to the graph in Fig.~\ref{1PI}$(b)$
including the 4-vertex are
\begin{eqnarray}
\langle\tilde\phi(p)\tilde\phi(-p)\rangle_1^{V_4}&=&
\left(\frac{2\pi}{\sqrt{\lambda}}\right)^2\frac{1}{(p^2+4)^2}
\int \frac{d^2q}{(2\pi)^2} \Bigg{[} -2\frac{q^2+4+p^2}{q^2+4}
-16\frac{q_1^2+1}{q^2+2}-10\frac{q^2}{q^2}
+32\frac{q_1^2+1}{q^2+1}\Bigg{]}\cr
&=&\left(\frac{2\pi}{\sqrt{\lambda}}\right)^2
\frac{1}{(p^2+4)^2}\int \frac{d^2q}{(2\pi)^2}\Bigg{[}
-2\frac{p^2}{q^2+4}+16\frac{1}{q^2+1}\Bigg{]}\,.
\label{phi-V4}
\end{eqnarray}
Here in the second step we have dropped purely power-like divergences (using rotational
invariance when needed). Finally, the tadpole diagram of Fig.~\ref{non-1PI} gives
\begin{eqnarray}
\langle\tilde\phi(p)\tilde\phi(-p)\rangle_1^{\rm tad.}=
\left(\frac{2\pi}{\sqrt{\lambda}}\right)^2\,8\frac{p_0^2-p_1^2}{(p^2+4)^2}
\int \frac{d^2q}{(2\pi)^2} \frac{q_1^2+1}{q^2+1}
=\left(\frac{2\pi}{\sqrt{\lambda}}\right)^2\, 4\frac{p_0^2-p_1^2}{(p^2+4)^2}
\int \frac{d^2q}{(2\pi)^2} \frac{1}{q^2+1}\,.
\label{phi-tad}
\end{eqnarray}
While the integrals appearing in the $\langle x(p)x^*(-p)\rangle$ 2-point function
could also be analyzed directly through a Feynman parametrization, this does not seem
to be convenient for the integrals appearing in eq. (\ref{phi-V3V3}). Tensor-reducing the numerator factors yields:
%
\begin{eqnarray}
&&\langle\tilde\phi(p)\tilde\phi(-p)\rangle_1^{V_3V_3}=\left(\frac{2\pi}{\sqrt{\lambda}}
\right)^2\frac{1}{(p^2+4)^2}\Bigg{[}
\frac{2}{\pfour}(p^2+4)^2(p_0^2-p_1^2)^2\,G[4,4;p^2]-8\,{\rI}[4]\cr
&&~~~~~~~~+\left((p^2+4)^2+64\left(\frac{p_1^4}{\pfour}-\frac{p_1^2}{p^2}\right)\right)\,G[2,2;p^2]
-2(4+p^2)\,{\rI}[2]\cr
&&~~~~~~~~-4\frac{p_1^2}{\pfour}(p^2+4)(\pfour+4p_1^2)G[1,1;p^2]+8p_1^2{\rI}[1]\Bigg{]}
\label{phi-V3-simp}
\end{eqnarray}
Summing up the contributions in
eq.~(\ref{phi-V4}),(\ref{phi-tad}) and (\ref{phi-V3-simp}) we
see that all UV divergences cancel out, since they combine
into $2(p^2+4)\left(2{\rI}[1]-{\rI}[2]-{\rI}[4]\right)=(p^2+4)\frac{3\ln 2}{2\pi}$. So the final result for the 1-loop correction to the $\tilde\phi$ propagator is
\begin{eqnarray}
&& \langle \tilde{\phi}(p)\tilde{\phi}(-p)\rangle_1 =\left(\frac{2\pi}{\sqrt{\lambda}}\right)^2\,\frac{1}{(p^2+4)^2}\Bigg{[}
\frac{2}{\pfour}(p^2+4)^2(p_0^2-p_1^2)^2\,G[4,4;p^2]\no \\
&&+\left((p^2+4)^2+64\left(\frac{p_1^4}{\pfour}-\frac{p_1^2}{p^2}\right)\right)\,G[2,2;p^2]
-4\frac{p_1^2}{\pfour}(p^2+4)(\pfour+4p_1^2)G[1,1;p^2]+(p^2+4)\frac{3\ln 2}{2\pi}\Bigg{]}\cr
&&\equiv \left(\frac{2\pi}{\sqrt{\lambda}}\right)^2\,\frac{1}{(p^2+4)^2} F^{(1)}_{\tilde{\phi}\tilde{\phi}}(p_0,p_1)
\label{phi-2pt}
\end{eqnarray}
As we will discuss in detail in section~4,
to obtain the corrected dispersion relation, it is sufficient as before to
evaluate $F^{(1)}_{\tilde{\phi}\tilde{\phi}}(p_0,p_1)$ on-shell, i.e. at $p^2=-4$.
 From \rf{mmm}  one finds that
\begin{equation}
G[4,4;p^2=-4]=\frac{1}{24\sqrt{3}}\ , \ \ \ \ \   \qquad G[2,2;p^2=-4]=\frac{1}{16}\,.
\end{equation}
The third integral,  $G[1,1;p^2=-4]$ diverges, since
\begin{equation}
G[1,1;p^2]=
\frac{{\rm arcsinh}\frac{|p|}{2}}{ \pi \ |p| \sqrt{p^2+4 }}\ .
\label{G11}
\end{equation}
This integral corresponds to the graph
in Fig.~\ref{1PI}$(a)$ with a fermion loop and
external $\tilde\phi$ field.
The divergence can be attributed
 to the fact that due to the value of the masses,
  the $m=2$ boson
 may  {\it kinematically} decay\foot{Note that $G[1,1;p^2]$ is the only potential source for an imaginary contribution (for $p^2\le -4$) to $F^{(1)}_{\tilde{\phi}\tilde{\phi}}(p_0,p_1)$.} into a pair of on-shell $m=1$ fermions.
However,  the integral $G[1,1;p^2]$ appears in our result (\ref{phi-2pt}) multiplied
 by a factor $(p^2+4)$, and hence
\begin{equation}
{\rm lim}_{p^2\rightarrow -4}\ \  (p^2+4)G[1,1;p^2]=0\ .
\end{equation}
Therefore, due to dynamical reasons (the particular structure of the relevant cubic coupling in
\rf{cubic}), the
 fermionic loop contribution actually drops out of  the on-shell
self-energy.
%
In section 4 we will see that, in contrast to the case considered in
\ci{zar}, the off-shell behavior of this term
does not affect the existence of a real simple pole of the quantum corrected
propagator due to the presence of other non-vanishing contributions.
Indeed, we  find that
\begin{equation}
F^{(1)}_{\tilde{\phi}\tilde{\phi}}(p_0,p_1)|_{p^2=-4} = \frac{1}{4}\,p_1^2(p_1^2+4)\,.
\label{Fphi}
\end{equation}
Then our final result is that the dispersion relation for $\tilde\phi$ at 1-loop is
\begin{equation}
p^2=-4+\frac{\pi}{2\sqrt{\lambda}}\, p_1^2(p_1^2+4)+{\cal O}(\frac{1}{\lambda})\,.\la{dri}
\end{equation}
In Minkowski signature this gives
\begin{equation}
\label{dre}
E^2=\left(p_1^2+4\right)\left[1-\frac{\pi}{2\sqrt{\lambda}}p_1^2\right]+{\cal O}(\frac{1}{\lambda})\,.
\end{equation}
We observe that the mass, defined as the energy at zero momentum,
 does not receive corrections at 1-loop order, unlike the  case of the
 $m=\sqrt{2}$ boson (cf. \rf{tamm}).


\subsection{The massless bosons}

To conclude the analysis of bosonic fluctuations, let us consider the 1-loop propagator for the 5 massless fields $y^a$. The contributions from the topology in Fig.~\ref{1PI}$(a)$ is
\begin{eqnarray}
&&\langle y^a(p) y^b(-p) \rangle_1^{\rm V_3 V_3} =
\left(\frac{2\pi}{\sqrt{\lambda}}\right)^2\, \frac{\delta^{ab}}{\pfour} \int
\frac{d^2q}{(2\pi)^2}\Bigg{[}\frac{4(p_0 q_0-p_1 q_1)^2}{q^2((p-q)^2+4)} \no \\
&&\ \  \ \ \ \ \ \ \ \ \ \ \ \   \ \ \ + \ 8\frac{((p_1-q_1)^2+1)(q_1^2+1-q_0 (p_0-q_0))}{(q^2+1)((p-q)^2+1)}\cr
&&\ \ \ \ \ \ \  \ \  \ \ \ \ \ \ \ +\ 4\frac{p_0^2 q_0(p_0-q_0)}{(q^2+1)((p-q)^2+1)}
-16\frac{(q_1^2+1)p_0(p_0-q_0)}{(q^2+1)((p-q)^2+1)}\Bigg{]}\,.
\end{eqnarray}
Here the first term in square brackets
comes from the purely bosonic contribution, while the other
 three are due to fermions. We can now simplify these
 integrals by performing a tensor reduction of the numerators.
  To highlight the fact that the bosonic contribution is IR finite,
  we can make use of the identity ($G[m_1^2,m_2^2;p^2]$ was defined in (\ref{GG}))
\begin{equation}
G[0,m^2;p^2]=\frac{1}{p^2+m^2}\left(\frac{1}{2\pi}\ln\frac{m^2+
p^2}{m^2}-{\rI}[m^2]+{\rI}[0]\right)\,.
\label{G01-identity}
\end{equation}
Then we have
\begin{eqnarray}
&&\langle y^a(p) y^b(-p) \rangle_1^{\rm V_3 V_3} = \left(\frac{2\pi}{\sqrt{\lambda}}
\right)^2 \frac{\delta^{ab}}{\pfour} \Bigg{[}\frac{1}{2\pi \pfour}(4+p^2)(\pfour-8 p_0^2 p_1^2)\ln(1+\frac{p^2}{4})+2p^2{\rm I}[4]\cr
&& ~~~~~~~~~~~~~~~~~~~~~
+ \frac{4}{p^2} p_0^2 (\pfour+4p_1^2)\,G[1,1;p^2]+(4-8p_0^2){\rI}[1]\Bigg{]}\,.
\end{eqnarray}
The contributions coming from the 4-vertex are
\begin{eqnarray}
&&\langle y^a(p) y^b(-p) \rangle_1^{\rm V_4}=\left(\frac{2\pi}{\sqrt{\lambda}}\right)^2 \frac{\delta^{ab}}{\pfour}
\int \frac{d^2q}{(2\pi)^2} \Bigg{[}\frac{5}{2}\frac{p^2}{q^2}-2\frac{p^2}{q^2+4}-8\frac{q_1^2+1}{q^2+1}\Bigg{]}\cr
&&~~~~~~~~~~~~~~~~~~
=\left(\frac{2\pi}{\sqrt{\lambda}}\right)^2 \frac{\delta^{ab}}{\pfour}\Bigg{[}\frac{5}{2}p^2{\rI}[0]-2p^2{\rI}[4]-4{\rI}[1]\Bigg{]}\,.
\end{eqnarray}
Finally, the contribution of the diagram involving the tadpole is
\begin{eqnarray}
&&\langle y^a(p) y^b(-p) \rangle_1^{\rm tad.}=\left(\frac{2\pi}{\sqrt{\lambda}}\right)^2 \frac{\delta^{ab}}{\pfour}
\int \frac{d^2q}{(2\pi)^2} 4\frac{p_0^2-p_1^2}{q^2+1}=
\left(\frac{2\pi}{\sqrt{\lambda}}\right)^2 \frac{\delta^{ab}}{\pfour} 4(p_0^2-p_1^2){\rI}[1]\,.
\end{eqnarray}
Summing everything up, we get
\begin{eqnarray}
&&\langle y^a(p) y^b(-p) \rangle_1 =\left(\frac{2\pi}{\sqrt{\lambda}}\right)^2 \frac{\delta^{ab}}{\pfour}
\Bigg{[} \frac{1}{2\pi \pfour}(4+p^2)(\pfour-8 p_0^2 p_1^2)\ln\left(1+\frac{p^2}{4}\right)\cr
&&~~~~~~+\frac{4}{p^2} p_0^2 (\pfour+4p_1^2)\,G[1,1;p^2]-4p^2{\rI}[1] +\frac{5}{2}p^2 {\rI}[0]\Bigg{]}\equiv \left(\frac{2\pi}{\sqrt{\lambda}}\right)^2 \frac{\delta^{ab}}{\pfour}F^{(1)}_{yy}(p_0,p_1)\,.
\label{xeyy}
\end{eqnarray}
To obtain the corrected dispersion relation, we
again
 need to evaluate $F^{(1)}_{yy}(p_0,p_1)$ on-shell, i.e.
  at $p^2=0$, or $p_0^2=-p_1^2$. Then all of the UV and IR
  divergences in the result drop out.\foot{IR divergences are nevertheless expected to show up at higher orders  or in 
 more complicated  objects like scattering amplitudes indicating that massless 2d 
 scalars  should  not be proper asymptotic states  at finite coupling.}
   Using that at small $p^2$ we have $G[1,1;p^2]=\frac{1}{4\pi}-\frac{p^2}{24\pi}+\ldots$, and taking
    the limit $p_0^2\rightarrow -p_1^2$ in the above expression, we obtain
\begin{equation}
F^{(1)}_{yy}(p_0,p_1){|_{p_0^2\rightarrow -p_1^2}}=\frac{7}{6\pi}p_1^4\,.
\label{Fy}
\end{equation}
Notice that in  the limit $p_0^2\rightarrow -p_1^2$
there is a cancellation of singular terms
between the bosonic and
fermionic contributions.
 Using this on-shell value, we conclude
that the 1-loop corrected dispersion relation for the 5 massless fields is
\begin{equation}
p^2=\frac{7}{3\sqrt{\lambda}}p_1^4+{\cal O}(\frac{1}{\lambda})\,.
\end{equation}
Going to Minkowski signature $(p_0,p_1)=(i E,p_1)$, this gives the dispersion relation
\begin{equation}
\label{drely}
E^2=p_1^2\left(1-\frac{7}{3\sqrt{\lambda}}p_1^2\right)+{\cal O}(\frac{1}{\lambda})\,.
\end{equation}

\subsection{The fermions}

Let us finally look at the 1-loop correction to the 2-point function of the 8 fermionic fluctuations. The intermediate expressions for the various diagrams are much more involved in this case, and therefore we will omit most of the details here.

From the explicit 1-loop calculation, one
can see that the  non-vanishing entries
 in  the fermion 2-point function matrix remain
  the same ones as at the tree level (\ref{fermions-tree}).
  In fact, the
   1-loop corrections to the fermionic propagators can be written in the form
\begin{eqnarray}
&&\langle\tilde\theta^i(p)\tilde\eta^j(-p)\rangle_1 =-\left(\frac{2\pi}{\sqrt{\lambda}}\right)^2\frac{p_1-
\mathrm{i}}{(p^2+1)^2}(\rho_6^{\dagger})^{ij}\, F^{(1)}_{\tilde\theta\tilde\eta}(p_0,p_1) \ ,\cr
&&
\langle\tilde\theta_i(p)\tilde\eta_j(-p)\rangle_1 =-\left(\frac{2\pi}{\sqrt{\lambda}}\right)^2\frac{p_1-
\mathrm{i}}{(p^2+1)^2}\rho^6_{ij}\, F^{(1)}_{\tilde\theta\tilde\eta}(p_0,p_1) \cr
&&\langle\tilde\theta^i(p)\tilde\theta_j(-p)\rangle_1=
-\left(\frac{2\pi}{\sqrt{\lambda}}\right)^2\frac{p_0}{(p^2+1)^2}\delta^i_j\,F^{(1)}_{
\tilde\theta\tilde\theta^{\dagger}}(p_0,p_1)\cr
&&\langle\tilde\eta^i(p)\tilde\eta_j(-p)\rangle_1=
-\left(\frac{2\pi}{\sqrt{\lambda}}\right)^2\frac{p_0}{(p^2+1)^2}\delta^i_j\,F^{(1)}_{
\tilde\eta\tilde\eta^{\dagger}}(p_0,p_1)\,.
\label{fermions-1loop}
\end{eqnarray}
Here $F^{(1)}_{\tilde\theta\tilde\eta}, F^{(1)}_{\tilde\theta\tilde\theta^{\dagger}}, F^{(1)}_{
\tilde\eta\tilde\eta^{\dagger}}$ are $SU(4)$ scalars  which  are independent functions
 when momenta are off-shell. However,  it turns out that they coincide up to
  terms which vanish on-shell (i.e. at $p^2=-1$). Computing the relevant
  1-loop diagrams, reducing them to scalar integrals and dropping terms
   which do not contribute on-shell, one indeed finds that
\begin{eqnarray}
&&F^{(1)}_{\tilde\theta\tilde\eta}(p_0,p_1)|_{_{p^2=-1}}
=F^{(1)}_{\tilde\theta\tilde\theta^{\dagger}}(p_0,p_1)|_{_{p^2=-1}}
=F^{(1)}_{\tilde\eta\tilde\eta^{\dagger}}(p_0,p_1)|_{_{p^2=-1}}\cr
&&
=16p_1^2(1+p_1^2)\,G[1,2;p^2]+4(1+p_1^2)
\left({\rI}[4]+2{\rI}[1]-\frac{5}{4}{\rI}[0]\right)\cr
&&~~+4(1+p_1^2)\left(\frac{5}{4}{\rI}[0]-{\rI}[4]\right)\cr
&&~~-8(1+p_1^2){\rI}[1]\,,
\end{eqnarray}
where the terms in the three lines correspond respectively
 to the contributions of the diagram in Fig.~\ref{1PI}$(a)$,  Fig.~\ref{1PI}$(b)$ and
Fig.~\ref{non-1PI}.
Notice that the
 integral $G[1,4;p^2]$ turns out to be absent when restricting
  to on-shell momenta, and so does the finite part of the
  integral $G[1,0;p^2]$ (to extract the finite part, we can
   use the identity (\ref{G01-identity})). Summing up the
   contributions in the three lines above, we find  that
   the  UV and IR divergences cancel on-shell.
   Using that $G[1,2;p^2=-1]=1/16$ we are then left with
\begin{eqnarray}
\label{Ff}
F^{(1)}_{\tilde\theta\tilde\eta}(p_0,p_1)|_{_{p^2=-1}}
=F^{(1)}_{\tilde\theta\tilde\theta^{\dagger}}(p_0,p_1)|_{_{p^2=-1}}
=F^{(1)}_{\tilde\eta\tilde\eta^{\dagger}}(p_0,p_1)|_{_{p^2=-1}}
=p_1^2(1+p_1^2)\,.
\end{eqnarray}
Inserting this into the 1-loop propagators in (\ref{fermions-1loop}) and
comparing to the tree level expressions (\ref{fermions-tree}), we see that
the effect of the 1-loop correction is to shift all denominators by the same amount
\begin{equation}
\frac{1}{p^2+1}\rightarrow \frac{1}{p^2+1-\frac{2\pi}{\sqrt{\lambda}}\,p_1^2(1+p_1^2)}\,,
\end{equation}
so that for all fermions we obtain the same dispersion relation (as expected by
the  $SO(6)$ symmetry)
\begin{equation}
p^2=-1+\frac{2\pi}{\sqrt{\lambda}}\,p_1^2(1+p_1^2)+{\cal O}(\frac{1}{\lambda})\,.
\end{equation}
In Minkowski signature 
this gives
\begin{equation}
\label{drelf}
E^2=\left(p_1^2+1\right)\left[1-\frac{2\pi}{\sqrt{\lambda}}\,p_1^2\right]
+{\cal O}(\frac{1}{\lambda})\,,
\end{equation}
which is again in agreement with the ABA result of \cite{bas}.


\section{Comments on poles of the  2-point function, physical states and stability}

In the previous section we evaluated the 1-loop correction to the 2-point function
of the worldsheet excitations around the  long spinning string or, equivalently,
the null cusp minimal surface. We carried out
the computation on a Euclidean worldsheet; upon analytic continuation to  2d Lorentzian
signature ($p_0 \to i E$),  the 1-loop corrections
modify the  dispersion relation of these excitation by shifting
position of the pole of the tree-level 2-point functions.
 In general, depending on the precise structure
of the 1-loop correction to the amputated 2-point function
(``self-energy'' operator)  $\Pi (p_0,p_1; \l)$, the
position of the real pole in the  classical propagator may get shifted
but it may also happen   that the pole  may disappear being replaced by a branch cut.
In the latter case  the corresponding field will no longer  represent an asymptotic particle
state, as  was suggested in \ci{zar} for the heavier BMN field
in string theory in $AdS_4\times \mathbb{CP}^3$.
As we have seen, in our case the situation  is different:
all classical excitations, including the heaviest  $AdS_3$ mode $\td \phi$,
continue to exhibit poles in their
quantum-corrected  propagator.
Let us discuss the difference with  the example  in \ci{zar}   in  more  detail.


\subsection{Existence of poles in the 2-point function}

Consider a field $\Phi$ with the classical mass-shell condition $p^2 +m^2 =0$
and the quantum-corrected  2-point function  having a generic form
($p^2 = p_0^2 + p_1^2 = - E^2 + p_1^2$)
\be
\langle\Phi(p)\Phi(-p)\rangle=\frac{1
}{p^2+m^2-\Pi (p_0,p_1; \l) }\ ,
\ \ \ \ \ \ \ \ \ \ \ \ \ \
\Pi (p_0,p_1; \l)= {\te  { 2 \pi \ov \sql }}   F^{(1)}(p_0, p_1) +  {\cal O}({\te { 1 \ov (\sql)^2}})  \ ,
\label{resg}
\ee
where  we assumed that the theory is such  that, like in
 \rf{cubic},\rf{quartic}, the    interaction terms
   do not in general preserve 2d Lorentz invariance.\foot{In our
 present case the 2d Lorentz invariance is broken ``spontaneously'' by the choice of the
 classical background and gauge-fixing condition.}
The  corrected dispersion relation is then determined by the real solutions of
$p^2+m^2-\Pi (p_0,p_1; \l)=0$. As usual,   the existence of a real simple pole
of the exact propagator
is equivalent to the possibility to identify the
 excitations of the field $\Phi$  as  free asymptotic particles.

Let us further assume  that  the 1-loop  term in  $\Pi$  has the
following  form
\be
F^{(1)}(p_0,p_1)=a_0 + a_{1/2} (p^2+m^2)^{1/2} + a_1  (p^2+m^2)  + a_2  (p^2+m^2)^2 \ , \ \ \ \ \ \ \
\ \ \     a_k = a_k (p_0, p_1)   \ ,
\label{gec}
\ee
where  the coefficients  $a_k$  should be   regular at $p^2=-m^2$.
This  is actually  the  structure we have   found
in \rf{phi-2pt} (and also in \rf{xexx} and \rf{xeyy} and the off-shell
fermion 2-point function;
 in these cases $a_{1/2}=0$).
 The non-analytic $ (p^2+m^2)^{1/2}$  term   appears in the propagator of
 the  $AdS_3$
 mode $\tilde\phi$ \foot{More precisely,
  the origin of this non-analyticity is due to the
   term $(p^2+4) G[1,1;p^2]\sim\sqrt{p^2+4}$ in \rf{phi-2pt}.}
   and also in the  case of the heavy BMN mode in \ci{zar}
 due to the possibility of a threshold
 two-body decay of the field  of mass $m$ into  two fields of mass $\ha m$
 present in the  theory.\foot{In \rf{gec}
 we  assumed that there is no $(p^2+m^2)^{-1/2}$
 term in $F^{(1)}$. Such term would appear  {\it e.g.} for a
 system of two fields  with masses $m$ and $ \ha m$ and a cubic interaction without
 derivatives, i.e.
 $L = \del_\a \Phi\del_\a \Phi + m^2 \Phi^2  +
 \del_\a \Psi \del_\a \Psi + {1 \ov 4} m^2 \Psi^2 +  h \Psi ^2 \Phi$. In our case, as well
 as in \cite{zar},
  this long distance singularity is softened by the
 special
 structure of the relevant part of 3-vertex coupling in the action.
 }

In the case considered  in \ci{zar} it was conjectured  that  due
 to underlying supersymmetry
the leading coefficient $a_0$
should  vanish\foot{It would be interesting to check this conjecture
by an explicit computation.
}
 so that the non-trivial part of
$F^{(1)}$ close to the original  mass shell is essentially determined  by the threshold
term,
\be
F^{(1)} (p_0,p_1)_{_{AdS_4\times \mathbb{CP}^3}}
 = a_{1/2} (p^2+m^2)^{1/2} +  {\cal O}(p^2+m^2)  \ , \ \ \ \ \ \ \ \ \ \ \
\ \ \     a_{1/2}   = - { 1 \ov 4 \pi} p^2_1    \ ,
\label{geca}
\ee
where the  value of $a_{1/2}$ was found \ci{zar}
by the explicit 1-loop  computation.\foot{Here
for simplicity  we identify our expansion parameter ${1 \ov \sql} $
with $ 1 \ov \sqrt{ 2 \l}$ in \ci{zar}. Also, in \ci{zar} one had  $m^2 =1$ for the
 heavier BMN mode. The relation between our notation and that  of \ci{zar}
 is as follows: $E \to \omega, \ \ p_1 \to p, \ \  \ e(p_1)\equiv \sqrt{p_1^2+ m^2}
 \to   E(p) $.}
In this case  the condition  for existence of a
 pole in the propagator  \rf{resg}  may be written as
\be
&&  - E^2  + p_1^2  + m^2  -
{ 2 \pi \ov \sql}  a_{1/2} ( - E^2  + p_1^2  + m^2   )^{1/2}  =  {\cal O}({\te { 1 \ov (\sql)^2}})
 \ , \la{zoo}
\ee
i.e.  for  $E\approx  e(p_1) +  {\cal O}({\te { 1 \ov \sql}}) $     where
$e(p_1) \equiv \sqrt{ p_1^2  + m^2}$ we get
\be
E - e(p_1)  + { 2 \pi\  {a_{1/2} } \ov  \sql\  \sqrt{2e(p_1)} }    \sqrt{ e(p_1) - E } \
  =  {\cal O}({\te { 1 \ov (\sql)^2}})
 \ . \la{zaa}
\ee
This equation does not have a    solution  $E= e(p_1) +
  {\cal O}({\te { 1 \ov (\sql)^2}})$  for a real pole   in $E$
   since  $a_{1/2}  < 0$.\foot{Note that the  validity of this  statement again
   rests on  the conjecture that 2-loop and higher corrections to
   $\Pi$  in this case vanish  on the tree level mass shell $E=e(p_1)$.
   Otherwise the $ {\cal O}({\te { 1 \ov (\sql)^2}})$ term in the r.h.s. of \rf{zoo}
   may   lead to a non-trivial solution  $E= e(p_1) +
 {\te { q \ov (\sql)^2}}$  with real $q$.}
 The resulting conclusion of \ci{zar} was that in such case the
 particle of  mass $m$ dissolves  in the continuum of 2-particle states  of field of mass
 $\ha m$, i.e. it does not exist as an asymptotic state  in the spectrum
 at finite $\lambda$.

 The  main  difference  of this  case  compared to  our analysis of the
  propagator of the  $\tilde \phi$ mode  with $m^2=4$ is that here  $a_0 \not=0$
 in  \rf{phi-2pt}, i.e. $F^{(1)}_{\td \phi \td \phi }|_{_{p^2=-m^2}}= a_0  \not=0$
 (see \rf{Fphi}). As a result,
 the real pole
 of the corrected propagator is readily found   and was
  already  given  in \rf{dre}.
 Indeed, we find  (cf. \rf{zoo},\rf{zaa})
 \be
 \la{hoh}
&& - E^2  + p_1^2  + m^2  -   { 2 \pi \ov \sql} a_0 =  {\cal O}({\te { 1 \ov (\sql)^2}})
  \ , \la{kp} \\
&&  E= e(p_1) - { \pi \ov \sql } { a_0 \ov e(p_1) }   +  {\cal O}({\te { 1 \ov (\sql)^2}})
=e(p_1)\Big[1 - { \pi \ov 4\sql }p^2_1 \Big]   +  {\cal O}({\te { 1 \ov (\sql)^2}})
  \ ,  \la{ku}
\ee
 which is equivalent to \rf{dre} for $a_0= { 1 \ov 4 } p^2_1 [e(p_1)]^2$ in \rf{Fphi}.
As a result, this heavy $AdS_3$ mode
remains an asymptotic state  in the spectrum.
The same  analysis applies also to all the
other excitations we discussed in section 3  (which also  have $a_0\not=0$ in their
self-energy operator)
leading to the  dispersion relations in (\ref{tamm}), (\ref{drely}) and (\ref{drelf}).

\def \aa {{\rm a}}

As a remark, let us however note that there is still a possible scenario in which
 a slightly modified version of the argument of \cite{zar} described above appears 
 to apply. Suppose that the denominator of the all-loop resummed 2-point function
  takes the form 
\begin{eqnarray}
&&p^2+m^2-\Pi (p_0,p_1; \l) = \Delta -\frac{2\pi}{\sql}\ \aa_{1/2} \ \Delta^{1/2} 
-\frac{2\pi}{\sql}\sum_{k \ge 1} \aa_k \ \Delta^k  \la{newr} \\
&&\Delta\equiv p^2+\tilde{m}^2(p_0,p_1;\sql) \ , \  \ \ \ \ \  \ \ \ 
\no
\tilde{m}^2(p_0,p_1;\sql)=m^2-\frac{2\pi}{\sql}\ \aa_0(p_0,p_1;\sql)\ ,\\
&& \aa_r\equiv  \aa_r(p_0,p_1;\sql)=a_r(p_0,p_1)+{\cal O}(\frac{1}{\sql})\,,\  \ \ \ \  \ \ \ \ \ r=0, 1/2, 1 , ...
\no
\end{eqnarray}
Note that this structure is,  in principle,  compatible with our 1-loop 
result for $\langle \tilde \phi(p) \tilde \phi(-p)\rangle$. Expanding
 in powers of $\frac{1}{\sqrt{\lambda}}$, one would find a pole to any 
 order in perturbation theory, in particular reproducing the 1-loop
  dispersion relation as described above. However, the zero of the 
  exact denominator at $p^2=-\tilde{m}^2$ does not yield a simple 
  pole of the 2-point function but just a branch cut due to the 
  square root term. Therefore, as in \cite{zar}, one may be led 
  to the conclusion that the particle of mass $m$ dissolves in
   the continuum of two-particle states. While this is, in 
   principle, a possibility we cannot rule out, we would like 
   to stress that it relies on very strong assumptions regarding
    the analytic structure of the higher loop corrections to 
    the 2-point function, which, in the absence of explicit 
    calculations, appear to be not very well motivated. It
     would be interesting to explore this further by 
     examining, {\it e.g.}, the 2-loop corrections to the propagator.


It is interesting to further discuss the consequences of the presence of the
 non-analytic  term
 \begin{equation}
 F^{(1)}_{\tilde\phi\tilde\phi}(p_0,p_1)=a_{1/2}(p^2+4)^{1/2}+\ldots \la{sqq}
\end{equation}
 in the 1-loop self-energy
for the field  $\tilde\phi$ at higher orders in  perturbation theory
in  $1 \ov \sql$.
Here the explicit value of $a_{1/2}$ can be read off from eq.~(\ref{phi-2pt})
and the
omitted terms are analytic close to the mass-shell.
When solving perturbatively for the poles of the quantum corrected 2-point function, one
would need to evaluate $F^{(1)}_{\tilde\phi\tilde\phi}(p_0,p_1)$ at the
position of the 1-loop
corrected pole. According to eq.~(\ref{dri}),\rf{kp}, this is given by
$p^2+4=\frac{\pi}{2\sqrt{\lambda}}p_1^2(p_1^2+4)\ge 0$,
i.e. the energy $E$ in \rf{ku}
is shifted below the threshold, $E- e(p_1) \leq 0$.
 Therefore,
 the 1-loop self-energy remains {\it real} on the 1-loop corrected mass shell
 which is  consistent  with our conclusion that $\td \phi$
 represents a stable asymptotic state in the spectrum
  (the presence of an  imaginary
part would imply a non-vanishing decay width,
 see the next subsection).

We notice, at the same time, that the presence
of the square root term in \rf{sqq}
 introduces a potential non-analyticity in the on-shell   value of the
self-energy  of $\tilde\phi$  considered
 as a function of $\sqrt{\lambda}$ and $p_1$. Indeed,
\begin{equation}
(p^2+4)^{1/2}|_{_{p^2+4=\frac{\pi}{2\sqrt{\lambda}}p_1^2(p_1^2+4)}}\ \sim \
\lambda^{-1/4}\sqrt{p_1^2(p_1^2+4)}\,.
\end{equation}
As a  result, the next to the leading correction in the dispersion relation
\rf{ku}  would actually be of order ${\te { 1 \ov \l^{3/4} }}$
instead of the  expected  ${\cal O}({\te { 1 \ov (\sql)^2}})$ one.
It is possible that this non-analytic behavior, which would   single
out $\tilde\phi$ as being
very different from the other fluctuation fields, may be compensated
by suitable contributions
to the 2-loop self-energy $F^{(2)}_{\tilde\phi\tilde\phi}(p_0,p_1)$.\footnote{This
could
happen, for example, if $F^{(2)}_{\tilde\phi\tilde\phi}(p_0,p_1)$
contains a non-analytic term
proportional to $(p^2+4)^{-1/2}$.} Once again, an explicit
2-loop calculation appears to be needed in order
to settle this question.

\subsection{On higher-order stability of the $m=2$ field}\label{stability}

The field ${\tilde \phi}$ is kinematically allowed to decay into a pair of fermions and
the relevant interaction terms are present  in the Lagrangian (\ref{Lagrangian}).
The existence of this decay becomes therefore a dynamical question.
As we have seen above, the poles of the 1-loop corrected 2-point function are real.
The absence of an imaginary part at this order implies that ${\tilde\phi}$ has
vanishing decay width  $\Gamma$
at the tree-level.
Indeed, the two  are  related via  the optical theorem
\be
\Gamma= \frac{1}{m}{\rm Im}\ \Pi  |_{_{p^2=-m^2}} \ , \ \la{opp}
\ee
where in the present case $m^2= 4$
and  $\Pi = { 2 \pi \ov \sql} F_{\td\phi\td\phi}^{(1)}(p_0,p_1) + ...$
is the sum of all quantum corrections to the amputated on-shell  2-point function.
The knowledge of $n$-loop corrections to $\Pi$ thus  allows  one to deduce
 the $(n-1)$-loop decay width.
Since, as was already  discussed,  the on-shell 1-loop 2-point function found in section 3.2
is real,   the total tree-level
 decay width vanishes.

\begin{figure}
\begin{center}
\includegraphics[width=140mm]{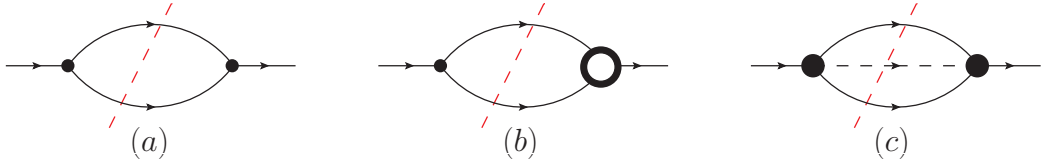}
\caption{Unitarity cuts at one and two  loops. In figure $(b)$ the right hand side part
 denotes a 1-loop vertex correction. The black dots
   in figure $(c)$ represent
on-shell 4-point amplitudes, which receive contributions from the 4-point vertices as
well as from the two three-point vertices.}
\label{cuts}
\end{center}
\end{figure}

Reversing the logic of the optical theorem and further using the generalized unitarity
to disentangle various multi-particle decay channels, we can gain information about the
1-loop stability of the ${\tilde \phi}$ field.
Indeed, the complete imaginary part may be evaluated as the on-shell limit of the
unitarity cut of the off-shell 2-point function. At the 1-loop level, see
Fig.~\ref{cuts}$(a)$,
this is nothing but the product
of certain on-shell tree-level vertices and two cut propagators (i.e. the residues of
Feynman propagators at the positive energy poles). At the 2-loop level,
 the unitarity cut
receives two types of contributions: two-particle cuts shown in  Fig.~\ref{cuts}$(b)$, and
three-particle cuts shown in Fig.~\ref{cuts}$(c)$. Through the generalized unitarity method
(see, {\it e.g.}, \ci{uni,alro} and references therein)
these two contributions may be computed and interpreted separately. Each of these
two contributions is responsible for one decay channel: Fig.~\ref{cuts}$(b)$ contains
contributions to a two-particle final state while Fig.~\ref{cuts}$(c)$ contains
contributions to a three-particle final state.

By analyzing the 1-loop unitarity cut it is easy to see that the product between the
${\tilde\phi}{\tilde\eta}{\tilde\theta}$  3-point vertex (the second line of
(\ref{cubic}))
and two-cut fermion propagators is proportional to $(p^2+4)^{1/2}$ where $p$ is the momentum
of ${\tilde\phi}$; it  thus vanishes for an on-shell field.
This shows that the appearance
of the similar factor arising from $(p^2+4)G[1,1;p^2]$ in $\langle{\tilde\phi}(p)
{\tilde \phi}(-p)\rangle$ does not rely on summing over all fermionic degrees of freedom.

This observation has an immediate consequence for the part of the cut of the 2-loop
2-point function that describes the decay of ${\tilde \phi}$ into two fermions,
Fig.~\ref{cuts}$(b)$. Indeed, this cut as well as the one in which the vertex correction
is
on the left side of
the cut are proportional to the same product of the
${\tilde\phi}{\tilde\eta}{\tilde\theta}$  3-point vertex and two-cut fermion propagators
and therefore vanish on shell. Consequently, the 1-loop partial decay width into two
fermions vanishes.\footnote{In drawing this conclusion we
 assumed that the 1-loop three-point
function is less singular than $(p^2+4)^{-1/2}$
 on the mass shell of ${\tilde\phi}$.}

The analysis of the three-particle
cut \ref{cuts}$(c)$ is more involved and will not be attempted here.
 It is possible that
 the most efficient way to extract it is from the 2-loop  expression  for the
  2-point
function which would be interesting to compute.

\section{Concluding remarks}

In this paper we used  superstring  sigma model perturbation theory to
compute the leading strong-coupling corrections to the dispersion relations of the
 fluctuation modes near  the long spinning string in $AdS_5$
 and compared the results with the asymptotic Bethe ansatz predictions \ci{bas}.

 There are several directions that may be
 interesting to explore.
 One is extending our computation  to the 2-loop level
 with a possibility  of further  comparison  to the corresponding terms in ABA \ci{bas}.
 That would  also check  our unitarity-based arguments about the stability
 of the heaviest $AdS_3$ mode $\tilde\phi$. Such 2-loop computation
 would be similar in spirit
 to a 3-loop computation of the cusp anomaly  and at the
 moment appears to be technically
 challenging.

 Also interesting would be the analysis of
 multi-excitations/bound states and the computation of
 the scattering matrix for all the fluctuations. Such a calculation
 would potentially access the higher-twist excitations discussed in \cite{bas}
and provide further detailed tests of the ABA.

Another  generalization is to the  case of non-zero angular momentum in
$S^5$, which should allow to determine
  corrections to the dispersion relations in
\rf{enn},\rf{ennn}. As discussed in the introduction, that may also
help clarify the fate of  the heaviest excitation mode.

Finally, it would be interesting to apply direct
string sigma model  techniques developed in \ci{us,uss} and here to
the computation of strong-coupling corrections to  the  null polygonal  Wilson loops
via the OPE approach of \ci{ald}.

\



{\bf Acknowledgments}\\
We would like to thank   F. Alday and J. Maldacena
for raising  the  interesting question about  superstring corrections
to the  excitation spectrum of the  long spinning string
that motivated this investigation. We also acknowledge helpful  discussions
with J. Collins, P. Vieira and D. Volin. 
We are grateful 
to B. Basso and  K. Zarembo for very useful comments on the first version  of this paper. 
This work was supported in part by the US National Science Foundation under
PHY-0855356 (R.Ro.), the US Department of
Energy under contracts DE-FG02-201390ER40577 (OJI) (R.Ro.) and the A.P. Sloan Foundation (R.Ro.). It was
also supported by EPSRC (R.Ri.). The work of S.G. is supported by Perimeter Institute
for Theoretical Physics. Research at Perimeter Institute is supported by the
Government of Canada through Industry Canada and by the Province of Ontario through
the Ministry of Research $\&$ Innovation.



\

\end{document}